\documentclass[a4paper]{report}
\usepackage[utf8]{inputenc}
\usepackage[T1]{fontenc}
\usepackage{RJournal}
\usepackage{amsmath,amssymb,array}
\usepackage{booktabs}

\usepackage{makecell}
\usepackage{verbatim,amssymb}
\usepackage{threeparttable} 
\usepackage{pifont}
\DeclareUnicodeCharacter{B0}{\degree}
\usepackage{amsmath,bm}
\usepackage{booktabs} 
\usepackage{array}
\usepackage{makecell}
\usepackage[skip=0.5\baselineskip]{caption}

\newcommand{\lsq}{\left[}
\newcommand{\rsq}{\right]}
\newcommand{\lbc}{\left \{ }
\newcommand{\rbc}{\right \} }

\newcommand{\cond}{{\, \vert \,}}
\newcommand{\cmark}{\ding{52}}%
\newcommand{\xmark}{\ding{53}}%

\newcommand{\YCF}{Y^{\text{CF}}}
\usepackage{mathrsfs}
\usepackage{gensymb}
\usepackage{enumerate}
\usepackage{algorithm,algpseudocode}

\begin{document}

\sectionhead{Contributed research article}
\volume{XX}
\volnumber{YY}
\year{20ZZ}
\month{AAAA}

\begin{article}
\title{CIMTx: An R Package for Causal Inference with Multiple Treatments using Observational Data}
\author{by Lianyuan Hu and Jiayi Ji}

\maketitle

\abstract{
  \CRANpkg{CIMTx} provides efficient and unified functions to implement modern methods for causal inferences with multiple treatments using observational data with a focus on binary outcomes. The methods include regression adjustment, inverse probability of treatment weighting, Bayesian additive regression trees, regression adjustment with multivariate spline of the generalized propensity score, vector matching and targeted maximum likelihood estimation. In addition, \CRANpkg{CIMTx} illustrates ways in which users can simulate data adhering to the complex data structures in the multiple treatment setting. Furthermore, the \CRANpkg{CIMTx} package offers a unique set of features to address the key causal  assumptions: positivity and ignorability. For the positivity assumption, \CRANpkg{CIMTx} demonstrates techniques to identify the common support region for retaining inferential units using inverse probability of treatment weighting, Bayesian additive regression trees and vector matching}. To handle the ignorability assumption,  \CRANpkg{CIMTx} provides a flexible Monte Carlo sensitivity analysis approach to evaluate how causal conclusions would be altered in response to different magnitude of departure from ignorable treatment assignment.

\hypertarget{introduction}{%
\section{Introduction}\label{introduction}}

Modern comparative effectiveness research (CER) questions often require comparing the effectiveness of multiple treatments on a binary outcome \citep{hu2020estimation}. To answer these CER questions, specialized causal inference methods are needed. Methods appropriate for drawing causal inferences about multiple treatments include regression adjustment (RA) \citep{rubin1973use,  linden2016estimating}, inverse probability of treatment weighting (IPTW) \citep{feng2012generalized,mccaffrey2013tutorial}, Bayesian Additive Regression Trees (BART) \citep{hill2011bayesian,hu2021variable, hu2020estimation}, regression adjustment with multivariate spline of the generalized propensity score (RAMS) \citep{hu2021estimation}, vector matching (VM) \citep{lopez2017estimation} and targeted maximum likelihood estimation (TMLE) \citep{rose2019double}.  Drawing causal inferences using observational data, however, inevitably requires assumptions. A key causal identification assumption is the \emph{positivity} or sufficient overlap assumption,  which implies that there are no values of pre-treatment covariates that could occur only among units receiving one of the treatments \citep{hu2020estimation}. Another key assumption requires appropriately conditioning on all pre-treatment variables that predict both treatment and outcome. The pre-treatment variables are known as confounders and this requirement is referred to as the \emph{ignorability} assumption (also as no unmeasured confounding) \citep{hu2022flexible}. An important strategy to handle the positivity assumption is to identify a common support region for retaining inferential units. The ignorability assumption can be violated in observational studies, and as a result can lead to biased treatment effect estimates. One widely recognized way to address such concerns is sensitivity analysis \citep{erik2007strengthening,hu2022flexible}.

The \CRANpkg{CIMTx} package provides a suite of functions to easily implement the causal estimation methods, many of which were recently developed \citep{lopez2017estimation,hu2020estimation,hu2021estimation}. In addition,  \pkg{CIMTx} provides strategies to define a common support region to address the positivity assumption using IPTW, BART, VM and implements a flexible Monte Carlo sensitivity analysis approach \citep{hu2022flexible} for unmeasured confounding to address the ignorability assumption.
Finally, \pkg{CIMTx} offers detailed examples of how to simulate data adhering to the complex structures in the multiple treatment setting. The simulated data can then be used by an analyst to compare the performance of different causal estimation methods. Table \ref{tab:methods-comparison} summarizes  key functionalities of \pkg{CIMTx} in comparison to recent R packages designed for causal inference with multiple treatments using observational data.  \pkg{CIMTx} provides a comprehensive set of functionalities: from simulating data to estimating the causal effects to addressing causal assumptions and elucidating their ramifications. 
To assist applied researchers and practitioners who work with observational data and wish to draw inferences about the effects of multiple treatments,  this article provides a comprehensive illustration of the \pkg{CIMTx} package. 


\begin{table}[h]
\caption{Comparisons of  R packages for causal inference.} \label{tab:methods-comparison}
\centering\label{tb:summary}
\setlength{\tabcolsep}{1pt}
\begin{tabular}{lcccccc} 
\toprule
\textbf{R packages} & \thead{Continuous \\ Outcome} & \thead{Binary \\ Outcome}  & \thead{Sensitivity \\ Analysis}   & \thead{Identification of\\ Common Support} & \thead{Design\\ factors}   & \thead{Estimation\\ procedure}  \\
\midrule
\CRANpkg{CIMTx} & \xmark  & \cmark & \cmark  &\cmark$^\ast$ &\cmark & \begin{tabular}{@{}c@{}}RA, IPTW-SL\\IPTW-Multinomial\\IPTW-GBM \\ VM, BART \\RAMS, TMLE\end{tabular}\\
\hline
\CRANpkg{PSweight} & \cmark   & \cmark & \xmark  &\cmark &\xmark& 
\begin{tabular}{@{}c@{}}OW, IPTW-SL\\IPTW-Multinomial\\IPTW-GBM \end{tabular}\\
\hline
\CRANpkg{twang}  & \cmark & \xmark & \cmark & \xmark &\xmark & IPTW-GBM\\
\hline
\CRANpkg{WeightIt} & \cmark &  \xmark & \xmark & \cmark &\xmark & \begin{tabular}{@{}c@{}}CBPS, IPTW-SL\\IPTW-Multinomial\\IPTW-GBM,EBCW\\IPTW-TSBW \end{tabular}\\
\hline
\CRANpkg{CBPS} & \cmark& \cmark  & \xmark & \xmark &\xmark & CBPS\\
\hline
\CRANpkg{optweight} & \cmark & \xmark  & \xmark & \xmark &\xmark & IPTW-TSBW \\

\bottomrule
\end{tabular}
\begin{tablenotes}
\footnotesize \item \cmark: the feature is offered in the method; \xmark\; indicates otherwise; RA: Regression adjustment; IPTW: Inverse probability of treatment weighting; BART: Bayesian additive regression trees; RAMS: Regression adjustment with multivariate spline of generalized propensity score; VM: Vector matching; TMLE: Targeted maximum likelihood estimation; CBPS: Covariate balancing propensity score; OW: Overlap weights; IPTW-Multinomial: Inverse probability of treatment weighting with weight estimated by multinomial logistic regression; IPTW-GBM: Inverse probability of treatment weighting with weight estimated by generalized boosted model; IPTW-SL: Inverse probability of treatment weighting with weight estimated by Super learner; IPTW-TSBW: Inverse probability of treatment weighting with targeted stable balancing weights; EBCW: Empirical balancing calibration weights.
\item $^\ast$: Identification of Common Support is only for VM, BART and IPTW related methods
\item References: \textbf{PSweight} (Version 1.1.4): \cite{PSweight};\textbf{twang} (Version 1.6) \cite{twang};\textbf{WeightIt}  (Version 0.10.2) \cite{WeightIt};\textbf{CBPS} (Version 0.22): \cite{CBPS};  \textbf{optweight} (Version 0.2.5): \cite{optweight};
\end{tablenotes}
\vspace{-0.05in}
\end{table}

\section{Design factors for data simulation}\label{sec:simulation}
\pkg{CIMTx} provides specific functions to simulate data possessing complex data characteristics of the multiple treatment setting. Seven design factors are considered: (1) sample size, (2) ratio of units across treatment groups,  (3) whether the treatment assignment model and the outcome generating model are linear or nonlinear,  (4) whether the covariates that best predict the treatment also predict the outcome well, (5) whether the response surfaces are parallel across treatment groups, (6) outcome prevalence, and (7) degree of covariate overlap. 

\subsection{Design factors (1)--(5)}\label{raio_of_units}
For the data generating process of treatment assignment, consider a multinomial logistic regression model, 
\begin{equation}\label{eq:trt_assign}
\begin{split}
\ln  \dfrac{P(W=1)}{P(W=T)} &= \delta_1 + \bm{X}\xi_1^L + \bm{Q}\xi_1^{NL}\\
\vdots \\
\ln  \dfrac{P(W=T-1)}{P(W=T)} &= \delta_{(T-1)} + \bm{X}{\xi_{(T-1)}^{L}} + \bm{Q}\xi_{(T-1)}^{NL}
\end{split}
\end{equation}
where $\bm{Q}$ denotes the nonlinear transformations and higher-order terms of the predictors $\bm{X}$. $\xi_1^L,\ldots,  \xi_{(T-1)}^{L}$ are vectors of coefficients for the untransformed versions of the predictors $\bm{X}$ and $\xi_1^{NL}, \ldots, \xi_{(T-1)}^{NL}$ for the transformed versions of the predictors captured in $\bm{Q}$. 
The intercepts $\delta_1, \ldots,\delta_{(T-1)}$  can be specified to create the corresponding ratio of units across $T$ treatment groups. The $T$ sets of potential response surfaces can be generated as follows:
\begin{equation} \label{eq:outcome_gen}
\begin{split}
E[Y(1) | \bm{X}]& = \text{logit}^{-1}  \{ \tau_1+  \bm{X}\gamma_1^{L} + \bm{Q} \gamma_1^{NL} \} \\
\vdots \\
E[Y(T) | \bm{X}]& =  \text{logit}^{-1}  \{ \tau_T+\bm{X}\gamma_T^{L} + \bm{Q} \gamma_T^{NL} \}
\end{split}
\end{equation}
where the coefficient setting $\gamma_1^L = \ldots = \gamma_T^L$, $\gamma_1^{NL} = \ldots = \gamma_T^{NL}$ and $\tau_1 \neq \ldots \neq \tau_T$ corresponds to the parallel response surfaces, and by assigning different values to $\gamma_w^L$ and $\gamma_w^{NL}$ and setting $\tau_1 = \ldots = \tau_T =0$, nonparallel response surfaces are generated, which imply treatment effect heterogeneity. Note that the predictors $\bm{X}$ and the transformed versions of the predictors $\bm{Q}$ in the treatment assignment model~\eqref{eq:trt_assign} can be different than those in the outcome generating model~\eqref{eq:outcome_gen} to create various degrees of alignment.  The observed outcomes are related to the potential outcomes through $Y_i = \sum_{w_i \in \mathscr{W}} Y_i(w)$. Covariates $\bm{X}$ can be generated from user-specified data distributions. 

\subsection{Outcome prevalence}\label{outcome_prevalence}
Values for parameters $\tau_1, \ldots, \tau_{T}$ in model~\eqref{eq:outcome_gen} can be chosen to create various outcome prevalence rates. The outcomes are considered rare if the prevalence rate is $< 5\%$. 

\subsection{Covariate overlap}\label{covariate_overlap}
With observational data, it is important to investigate how the sparsity of covariate overlap impacts the estimation of causal effects. We can modify the formulation of the treatment assignment model~\eqref{eq:trt_assign} to adjust the sparsity of overlap by including a multiplier parameter $\psi$ \citep{hu2021estimatingsim}  as follows:  
\begin{equation}\label{eq:trt_assign_overlap}
\begin{split}
\ln  \dfrac{P(W=1)}{P(W=T)} &= \delta_1 + \bm{X}\psi\xi_1^L + \bm{Q}\psi\xi_1^{NL}\\
\vdots \\
\ln  \dfrac{P(W=T-1)}{P(W=T)} &= \delta_{(T-1)} + \bm{X}\psi\xi_{(T-1)}^{NL} + \bm{Q}\psi\xi_{(T-1)}^{NL},
\end{split}
\end{equation}
where larger values of $\psi$ correspond to increased sparsity degrees of overlap. 

\subsection{Implementation in CIMTx}\label{sec:example_data_simulation}

We will first demonstrate the functionality of \code{data\_sim()} in \pkg{CIMTx} to simulate data in the multiple treatment setting using the above 7 design factors. We first use the \code{data\_sim} function to simulate a dataset with the following characteristics: (1) sample size = 500, (2) ratio of units  = 1:1:1 across three treatment groups,  (3) nonlinear treatment assignment and outcome generating models,  (4) different predictors for the treatment assignment and outcome generating mechanisms, (5) parallel response surfaces, (6) outcome prevalence = $(0.16, 0.51, 0.75)$ in three treatment groups  with an overall rate around 0.5 and (7) moderate covariate overlap. Note that for the design factor (6),  we can adjust \code{tau} to generate rare outcome events. 

The outputs of the simulated data object are: (1) \code{data\$covariates} for $\bm{X}$, (2) \code{data\$w} for  treatment indicators, (3) \code{data\$y} for observed binary outcomes,  (4) \code{data\$y\_prev} for the outcome prevalence rates,   (5) \code{data\$ratio\_of\_units} for the proportions of units in each treatment group,  (6) \code{data\$overlap\_fig} for the visualization of covariate overlap via boxplots of the distributions of true generalized propensity score (GPS).

\begin{Schunk}
\begin{Sinput}
library(CIMTx)
set.seed(1)
data <- data_sim(
  sample_size = 500, n_trt = 3,
  x = c("rnorm(0, 0.5)",  # x1
        "rbeta(2, .4)",   # x2
        "runif(0, 0.5)",  # x3
        "rweibull(1, 2)", # x4
        "rbinom(1, .4)"),   # x5
  # linear terms in parallel response surfaces
  lp_y = rep(".2*x1 + .3*x2 - .1*x3 - .1*x4 - .2*x5", 3), 
  # nonlinear terms in parallel response surfaces
  nlp_y  = rep(".7*x1*x1  - .1*x2*x3", 3), 
  align = F,# different predictors used in treatment and outcome models
  # linear terms in treatment assignment model
  lp_w = c(".4*x1 + .1*x2  - .1*x4 + .1*x5",   # w = 1
           ".2*x1 + .2*x2  - .2*x4 - .3*x5"),  # w = 2
  # nonlinear terms in treatment assignment model
  nlp_w = c("-.5*x1*x4  - .1*x2*x5", # w = 1
            "-.3*x1*x4 + .2*x2*x5"), # w = 2
  tau = c(-1.5, 0, 1.5), delta = c(0.5, 0.5), psi = 1)
\end{Sinput}
\end{Schunk}

In this simulated dataset, the ratio of units (\code{data\$ratio\_of\_units}) and outcome prevalences (\code{data\$y\_prev}) are:

\begin{Schunk}
\begin{Soutput}
#> w
#>    1    2    3 
#> 0.35 0.35 0.30 
\end{Soutput}
\end{Schunk}


\begin{Schunk}
\begin{Soutput}
#>         w y_prev
#> 1       1   0.16
#> 2       2   0.51
#> 3       3   0.75
#> 4 Overall   0.46
\end{Soutput}
\end{Schunk}

\begin{figure}[H]
\centering
\includegraphics[width = 0.95\textwidth]{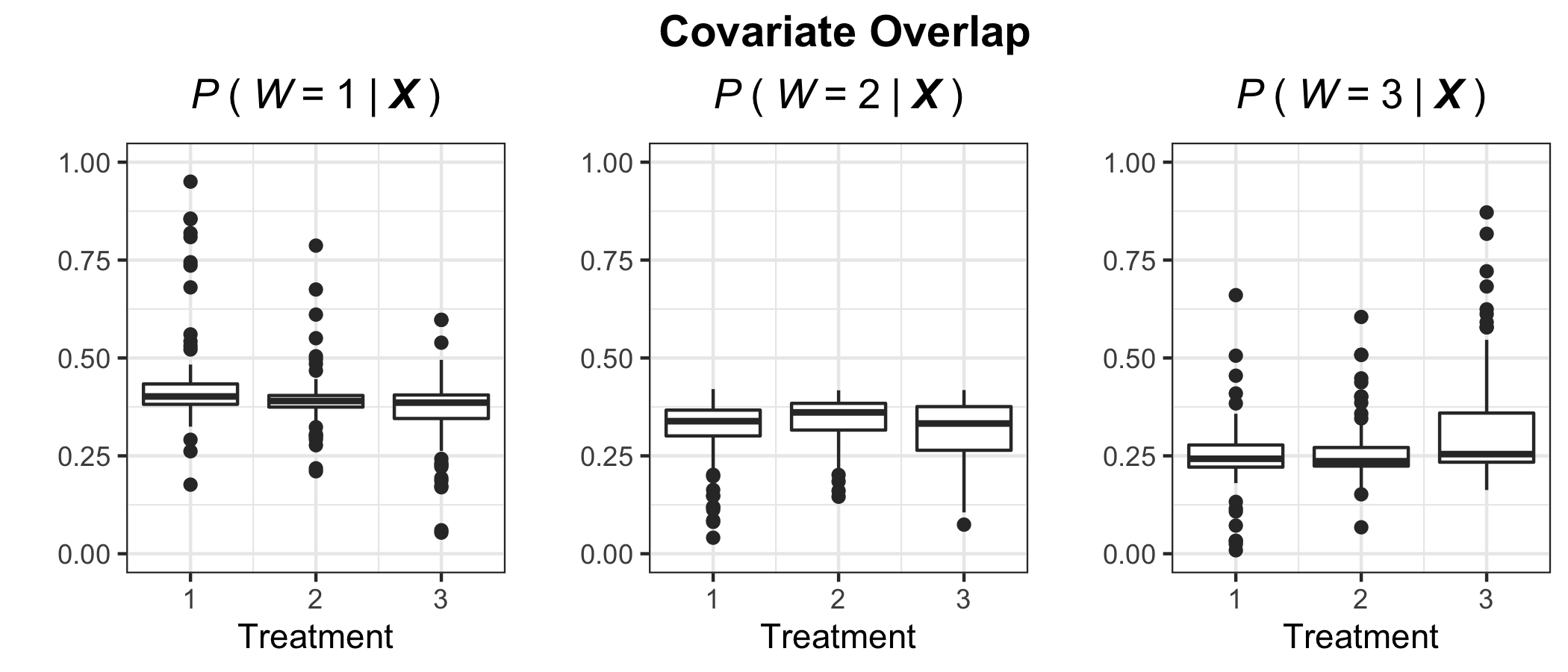}
\caption{Moderate overlap with \code{psi = 1}. Each panel presents boxplots by treatment group of the true generalized propensity score for one of the treatments  for every unit in the sample. }
\label{fig:covariate_overlap_moderate_plot}
\end{figure}
\begin{figure}[H]
\centering
\includegraphics[width = 0.95\textwidth]{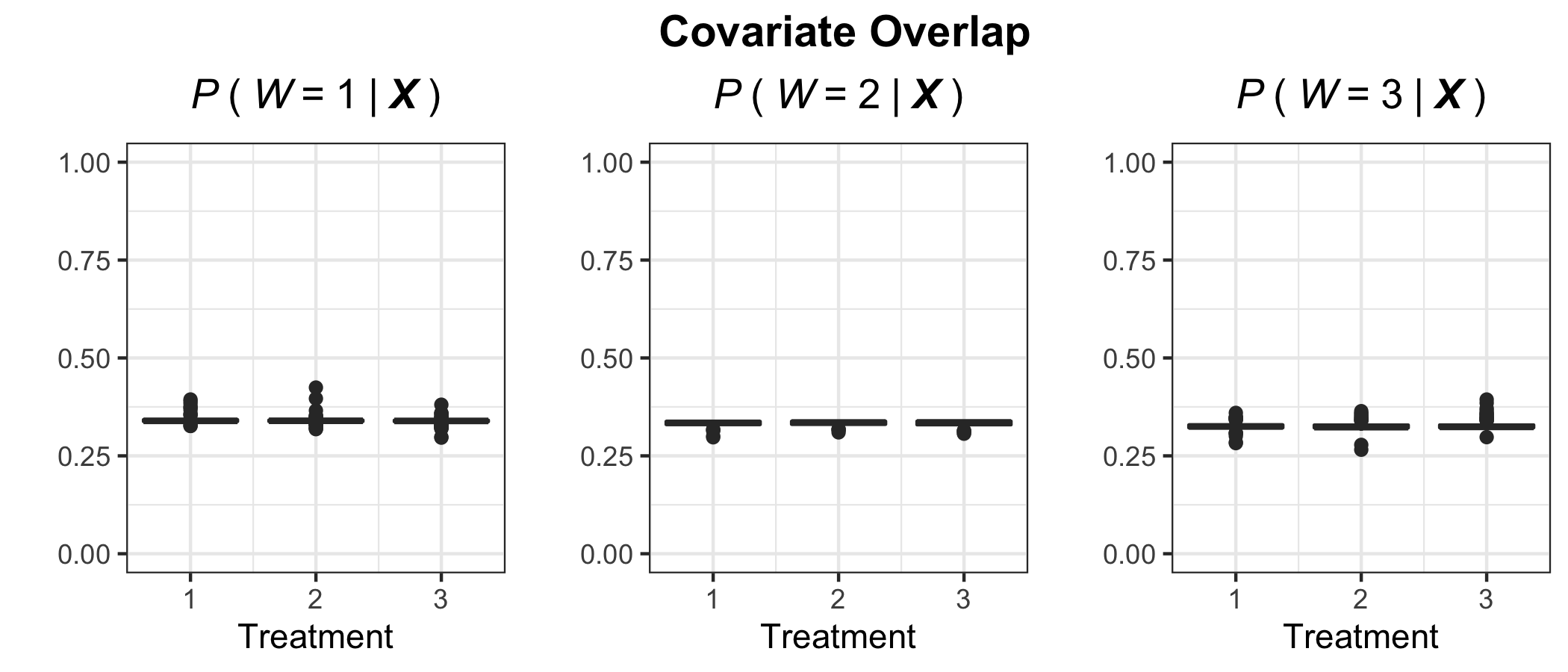}
\caption{Strong overlap with \code{psi = 0.1}. Each panel presents boxplots by treatment group of the true generalized propensity score for one of the treatments  for every unit in the sample. }
\label{fig:covariate_overlap_strong_plot}
\end{figure}

Figure \ref{fig:covariate_overlap_moderate_plot} (\code{data\$overlap\_fig}) shows the distributions of true GPS for each treatment group, suggesting moderate covariate overlap. We can change structures of the simulated data by modifying  arguments of the \code{data\_sim} function.  For example, setting \code{delta = c(1.5, 0.5)} yields unequal sample sizes across treatment groups with the ratio of unit $.6:.2:.2$. Assigning smaller values to \code{psi} can increase overlap: \code{psi = 0.1} corresponds to a strong covariate overlap as shown in Figure \ref{fig:covariate_overlap_strong_plot}. 

\section{Methodology and Implementation in CIMTx} \label{sec:causal_multiple}
\subsection{Estimation of causal effects} \label{sec:estimation}

Consider an observational study with $N$ individuals, indexed by $i =1, \ldots, N$, drawn randomly from a target population. Each individual was exposed to one and only one treatment, indexed by $W$. The goal of this study is to estimate the causal effect of treatment $W$ on a binary outcome $Y$. There are a total of $T$ possible treatments, and $W_i = w$ if individual $i$ is observed under treatment $w$, where $w \in \mathscr{W} = \{1, 2, \ldots, T\}$.  Pre-treatment measured confounders are indexed by $\bm{X}_i$. Under the potential outcomes framework,  \citep{rubin1974estimating, holland1986statistics}, individual $i$ has $T$ potential outcomes $\{Y_i(1), \ldots, Y_i(T)\}$ under each treatment of $\mathscr{W}$.  For each individual, at most one of the potential outcomes is observed -- the one corresponding to the treatment to which the individual is exposed. All other potential outcomes are missing, which is known as the fundamental problem of causal inference \citep{holland1986statistics}. In general,  three standard causal identification assumptions \citep{rubin1980randomization, hu2020estimation} need to be maintained in order to estimate the causal effects from observational data:  
\begin{enumerate}
    \item [(A1)] The stable unit treatment value assumption: there is no interference between units and there are no different versions of a treatment. 
    \item [(A2)] Positivity: the GPS for treatment assignment $e(\bm{X}_i)=P(W_i=1 \mid \bm{X}_i)$ is bounded away from 0 and 1. 
    \item [(A3)] Ignorability: pre-treatment covariates $\bm{X}_i$ are sufficiently predictive of both treatment assignment and outcome, $p(W_i \mid Y_i(1), \ldots, Y_i(T), \bm{X}_i) = p (W_i \mid \bm{X}_i)$.
\end{enumerate}
 The \pkg{CIMTx} package addresses assumption A(2) in the section of ``Identification of a common support region'' and A(3) in the section of ``Sensitivity analysis for unmeasured confounding''. 
 
Causal effects can be estimated by summarizing functionals of individual-level potential outcomes. For dichotomous outcomes, causal estimands can be the risk difference (RD), odds ratio (OR) or relative risk (RR).  For purposes of illustration, we define causal effects based on the RD.  Let $s_1$ and $s_2$ be two subgroups of treatments such that $s_1,s_2 \subset \mathscr{W}$ and $s_1 \cap s_2 = \emptyset$,  and define $|s_1|$ as the cardinality of $s_1$ and $|s_2|$ of $s_2$.  Two commonly used causal estimands are the average treatment effect (ATE), $ATE_{s_1,s_2}$, and the average treatment effect on the treated (ATT), for example, among those receiving $s_1$, $ATT_{s_1|s_1,s_2}$. They are defined as: 
\begin{equation}
\begin{split}
\label{eq:pop_est}
ATE_{s_1,s_2} &= E \bigg{[} \frac{\sum_{w \in s_1} Y_i(w)}{|s_1|} - \frac{\sum_{w' \in s_2} Y_i(w')}{|s_2|} \bigg{]},\\
ATT_{s_1|s_1,s_2} &= E \bigg{[} \frac{\sum_{w \in s_1} Y_i(w)}{|s_1|} - \frac{\sum_{w' \in s_2} Y_i(w')}{|s_2|} \bigg{|} W_i \in s_1 \bigg{]}.
\end{split}
\end{equation}

We now introduce six methods implemented in \pkg{CIMTx} for estimating the causal effects of multiple treatments: RA, IPTW, BART, RAMS, VM and TMLE.  

\paragraph{Regression adjustment} 
Regression adjustment \citep{rubin1973use,  linden2016estimating}, also known as model-based imputation \citep{imbens2015causal},  uses a regression model to impute missing potential outcomes: what would have happened to a specific individual had this individual received a treatment to which they were not exposed. RA regresses the outcomes on treatment and confounders, 
\begin{equation} \label{eq:RA_mod}
f(w,\bm{X}_i) = E[Y_i \cond  W_i =w, \bm{X}_i] = \text{logit}^{-1} \lbc \beta_0 + \beta_1 w + \bm{\beta}_2^{\top}\bm{X}_i\rbc,
\end{equation} 
where $\beta_0$ is the intercept, $\beta_1$ is the coefficient for treatment and $\bm{\beta}_2$ is a vector of coefficients for covariates $\bm{X}_i$. 
From the fitted regression model~\eqref{eq:RA_mod}, the missing potential outcomes for each individual are imputed using the observed data. The causal effects can be estimated by contrasting the imputed potential outcomes between treatment groups.  \pkg{CIMTx} implements RA with the Bayesian logistic regression model via the \code{bayesglm} function of the \CRANpkg{arm} package. For the ATE effects, we first average the $L$ predictive posterior draws $\{f^l(w,\bm{X}_i), l =1,\ldots, L\}$ over the empirical distribution of $\{\bm{X}_i\}_{i=1}^N$, and for the ATT effects using $s_1$ as the reference group, over the empirical distribution of $\{\bm{X}_i\}_{i: W_i \in s_1}$. We then take the difference of the averaged values between two treatment groups $w \in s_1$ and $w' \in s_2$. Inferences about treatment effect can be obtained based on the $L$ posterior average treatment effects. The 95\% credible interval is calculated using the 2.5th percentile and the 97.5th percentile of the posterior draws \citep{kruschke2014doing}.

In our package \pkg{CIMTx}, we can specify \code{method = "RA"} and \code{estimand = "ATE"} in the \code{ce\_estimate()} function to get the ATE effects via RA:

\begin{Schunk}
\begin{Sinput}
ra_ate_res <- ce_estimate(y = data$y, x = data$covariates, w = data$w, 
                          method = "RA", estimand = "ATE", ndpost = 100)
\end{Sinput}
\end{Schunk}

The estimates, standard errors and 95\% confidence intervals for the causal estimands would be printed using the \code{summary} generic function:

\begin{Schunk}
\begin{Sinput}
summary(ra_ate_res)
\end{Sinput}
\begin{Soutput}
#> $ATE12
#>      EST   SE LOWER UPPER
#> RD -0.28 0.04 -0.35 -0.21
#> RR  0.39 0.07  0.27  0.54
#> OR  0.26 0.06  0.17  0.40

#> $ATE13
#>      EST   SE LOWER UPPER
#> RD -0.60 0.06 -0.69 -0.47
#> RR  0.23 0.05  0.15  0.34
#> OR  0.07 0.02  0.03  0.13

#> $ATE23
#>      EST   SE LOWER UPPER
#> RD -0.31 0.04 -0.39 -0.25
#> RR  0.60 0.04  0.52  0.67
#> OR  0.25 0.05  0.17  0.34
\end{Soutput}
\end{Schunk}

Specifying \code{estimand = "ATT"} and setting \code{reference\_trt} will get us the ATT effects: 

\begin{Schunk}
\begin{Sinput}
ra_att_res <- ce_estimate(y = data$y, x = data$covariates,w = data$w, method = "RA", 
                          estimand = "ATT", ndpost = 100, reference_trt = 1)
summary(ra_att_res)
\end{Sinput}
\begin{Soutput}
#> $ATT12
#>      EST   SE LOWER UPPER
#> RD -0.28 0.05 -0.37 -0.18
#> RR  0.40 0.09  0.25  0.57
#> OR  0.27 0.08  0.16  0.44

#> $ATT13
#>      EST   SE LOWER UPPER
#> RD -0.59 0.06 -0.67 -0.46
#> RR  0.24 0.06  0.14  0.38
#> OR  0.07 0.03  0.03  0.13
\end{Soutput}
\end{Schunk}

\paragraph{Inverse probability of treatment weighting}

The idea of IPTW  was originally introduced by \cite{horvitz1952generalization} in survey research to adjust for imbalances in sampling pools. Weighting methods have been extended to estimate the causal effect of a binary treatment in observational studies, and more recently reformulated to accommodate multiple treatments \citep{imbens2000role, feng2012generalized,mccaffrey2013tutorial}. When interest is in estimating the pairwise ATE for treatment groups $s_1$ and $s_2$, a consistent estimator of  $ATE_{s_1,s_2}$ is given by the weighted mean, 
\begin{equation}
\widehat{ATE}_{s_1,s_2}=\frac{\sum_{i=1}^{N}Y_{i}I(W_{i}\in s_1)/|s_1|}{\sum_{i=1}^N  I(W_{i}\in s_1)r(W_i,\bm{X_{i}})} -\frac{\sum_{i=1}^{N}Y_{i}I(W_{i}\in s_2)/|s_2|}{\sum_{i=1}^N  I(W_{i}\in s_2)r(W_i,\bm{X_{i}})}
\end{equation}

where $r(w,\bm{X}_i)$ is the weights satisfying $r(w, \bm{X}_i) = 1 /P(W_i = w \mid \bm{X}_i )$, and $I(\cdot)$ is the indicator function. The \pkg{CIMTx} package provides three ways in which the weights can be estimated: (i) multinomial logistic regression \citep{feng2012generalized}, (ii) generalized boosted model (GBM) \citep{mccaffrey2013tutorial}, and (iii) super learner \citep{van2007super}. A challenge with IPTW is low GPS can result in extreme weights, which may yield erratic causal estimates with large sample variances \citep{little1988missing, kang2007demystifying}. This issue is increasingly likely as the number of treatments increases. Weight trimming or truncation can alleviate the issue of extreme weights \citep{cole2008constructing, lee2011weight}). \pkg{CIMTx} provides an argument for users to choose the percentile at which the weights should be truncated. We briefly describe the three weight estimators. 
\begin{enumerate}
    \item [(i)] The multinomial logistic regression model for treatment assignment is as follows:
\begin{align*}
P(W_i=w|\bm{X}_i) & =\frac{e^{\bm{\alpha'}_w\bm{X}_i}}{1+e^{\bm{\alpha'}_1\bm{X}_{i}} + \ldots +  e^{\bm{\alpha'}_{T-1}\bm{X}_{i}}}, 
\end{align*}
where $\bm{\alpha}_w$ is a vector of coefficients for $\bm{X}_i$ corresponding to treatment $w$, and can be estimated by using an iterative procedure such as generalized iterative scaling or iteratively reweighted least squares. 
\item[(ii)] GBM uses machine learning to flexibly model the relationships between treatment assignment and covariates. It does this by growing a series of boosted classification trees to minimize an exponential loss function. This process is  effective for fitting nonlinear treatment models characterized by curves
and interactions. The procedure of estimating the GPS can be tuned to find the GPS model producing the best covariate balance between treatment groups. 
\item[(iii)] Super learner is an algorithm that creates the optimally weighted average of several machine learning models.  The machine learning models can be specified via the \code{SL.library} argument of the \CRANpkg{SuperLearner} package. 
This approach has been proven to be asymptotically as accurate as the best possible prediction algorithm that is included in the library \citep{van2007super}. 
\end{enumerate}

IPTW can be implemented in \pkg{CIMTx} by setting a specific method and estimand. For IPTW estimators, variance can be estimated via a robust sandwich‐type variance estimator or a bootstrap variance estimator. In practice, a bootstrap variance estimator is often recommended. \citep{austin2016variance}. The following shows the code to estimate ATE using IPTW with weights estimated by multinomial logistic regression. 

\begin{Schunk}
\begin{Sinput}
iptw_multi_res <- ce_estimate(y = data$y, x = data$covariates , w = data$w, 
                              method = "IPTW-Multinomial", estimand = "ATE")
\end{Sinput}
\end{Schunk}

We can estimate the ATE effects with weights estimated by super learner and GBM by changing the argument of \code{method} to \code{"IPTW-SL",  "IPTW-GBM"} respectively. We can then estimate the causal effects and  bootstrap confidence intervals by setting \code{boot = TRUE}.

\begin{Schunk}
\begin{Sinput}
iptw_sl_trim_ate_res <- ce_estimate(y = data$y, x = data$covariates , w = data$w, 
                                    method = "IPTW-SL", estimand = "ATE", 
                                    sl_library =  c("SL.glm", "SL.glmnet", "SL.rpart"),
                                    trim_perc = c(0.05,0.95), boot = TRUE, 
                                    nboots = 100, verbose_boot = F)
\end{Sinput}
\begin{Soutput}
summary(iptw_sl_trim_ate_res)
#> $ATE12
#>      EST   SE LOWER UPPER
#> RD -0.34 0.05 -0.42 -0.24
#> RR  0.33 0.07  0.19  0.48
#> OR  0.20 0.06  0.10  0.33

#> $ATE13
#>      EST   SE LOWER UPPER
#> RD -0.59 0.05 -0.67 -0.46
#> RR  0.22 0.05  0.13  0.34
#> OR  0.07 0.02  0.04  0.13

#> $ATE23
#>      EST   SE LOWER UPPER
#> RD -0.25 0.05 -0.34 -0.15
#> RR  0.67 0.06  0.57  0.79
#> OR  0.34 0.09  0.21  0.54
\end{Soutput}
\end{Schunk}

\paragraph{Bayesian additive regression trees}
BART (\cite{chipman2010bart}) is a likelihood-based machine learning model and has been adapted into causal inference settings in recent years \citep{hill2011bayesian, hu2020estimation,hu2021estimation,hu2021estimatingsim,hu2021estimating}.  For a binary outcome, 
BART uses the probit regression
\begin{equation}
f(w,\bm{X}_i) = E[Y_i | W_i = w, \bm{X}_i] = \Phi \bigg{\{} \sum_{j=1}^J g_j(w, \bm{X}_i; T_j, M_j) \bigg{\}},
\end{equation}
where $\Phi$ is the the standard normal cumulative distribution function,  $(T_j, M_j)$ indexes a single subtree model in which $T_j$ denotes the regression tree and $M_j$ is a set of parameter values associated with the terminal nodes of the $j$th regression tree, $g_j(w,\bm{X}_i; T_j, M_j)$ represents the mean assigned to the node in the $j$th regression tree associated with covariate value $\bm{X}_i$ and treatment level $w$, and the number of regression trees $J$ is considered to be fixed and known. BART uses regularizing priors for $(T_j, M_j)$  to keep the impact of each tree small. 
Although the prior distributions can be specified via the \code{ce\_estimate} function of \pkg{CIMTx}, the default priors tend to work well and require little modification in many situations \citep{hill2011bayesian,hu2020estimation,hu2020tree}.  The details of prior specification and Bayesian backfitting algorithm for posterior sampling can be found in \citet{chipman2010bart}. The posterior inferences about the treatment effects can be drawn in a similar way as described in the Regression adjustment section. 

Setting \code{method = "BART"} and specifiying the \code{estimand = "ATE"} or \code{estimand = "ATT"} of the \code{ce\_estimate()} function implements the BART method. 

\begin{Schunk}
\begin{Sinput}
bart_res <- ce_estimate(y = data$y, x = data$covariates, w = data$w, method = "BART", 
                        estimand = "ATT", ndpost=100, reference_trt = 1)
\end{Sinput}
\end{Schunk}

\begin{Schunk}
\begin{Sinput}
summary(bart_res)
\end{Sinput}
\begin{Soutput}
#> $ATT12
#>      EST   SE LOWER UPPER
#> RD -0.38 0.07 -0.51 -0.25
#> RR  0.47 0.08  0.31  0.61
#> OR  0.21 0.07  0.10  0.35

#> $ATT13
#>      EST   SE LOWER UPPER
#> RD -0.56 0.07 -0.69 -0.43
#> RR  0.38 0.07  0.24  0.50
#> OR  0.06 0.03  0.02  0.13
\end{Soutput}
\end{Schunk}

\paragraph{Regression adjustment with multivariate spline of GPS} 
For a binary outcome, the number of outcome events can be small. The estimation of causal effects is challenging with rare outcomes because the great majority of units contribute no information to explaining the variability attributable to the differential treatment regimens in the health events \citep{hu2021estimation}. \cite{franklin2017comparing} found that regression adjustment on propensity score using one nonlinear spline performed best with respect to bias and root-mean-squared-error in estimating treatment effects. \cite{hu2021estimation} proposed RAMS, which accommodates multiple treatments by using a nonlinear spline model for the outcome that is additive in the treatment and multivariate spline function of the GPS as the following: 
\begin{equation}
f(W_i,\bm{X}_i) =  E[Y_i | W_i, \bm{X}_i]  =  \text{logit}^{-1} \bigg{\{}  \bm{\beta} W_i + h(\bm{R}(\bm{X}_i),\bm{\phi}) \bigg{\}},
\end{equation}
where $h(\cdot)$ is a spline function of the GPS indexed by $\bm{\phi}$ and $\bm{\beta} = [\beta_1, \ldots, \beta_T]^\top$ are regression coefficients associated with the treatment $W_i$. The dimension of the spline function $h(\cdot)$ depends on the number of treatments $T$. Confidence intervals of treatment effect estimates can be obtained using nonparametric bootstrap for RAMS \citep{hu2021estimation}.

In \pkg{CIMTx}, RAMS is implemented using the \code{gam()} function with tensor product smoother \code{te()} between treatments from the \CRANpkg{mgcv} package. Treatment effects can then be estimated by averaging and contrasting the predicted $\hat{f}(w,\bm{X}_i)$  between treatment groups. The RAMS can be called by setting \code{method = "RAMS-Multinomial"} and specifying the estimand \code{estimand = "ATE"} or \code{estimand = "ATT"}. 

\begin{Schunk}
\begin{Sinput}
rams_multi_res <- ce_estimate(y = data$y, x = data$covariates, w = data$w,
                              method = "RAMS-Multinomial", estimand = "ATE", 
                              boot = TRUE, nboots = 100, verbose_boot = F)

\end{Sinput}
\end{Schunk}

\paragraph{Vector matching}

\cite{lopez2017estimation} proposed the VM algorithm, which matches individuals with similar vector of the GPS.  VM obtains matched sets using a combination of $k$-means clustering and one-to-one matching with replacement within each cluster strata. Currently, VM is only designed to estimate the ATT effects. In  \pkg{CIMTx} , VM is implemented via  \code{method = "VM"}. The \pkg{CIMTx} does not provide confidence intervals for treatment effect estimates because the authors of this method,  \cite{lopez2017estimation}, did not provide an approach to estimate the sampling variance of the VM estimator.

To implement VM in \pkg{CIMTx},  we set the reference group  \code{reference\_trt = 1}, the number of clusters to form using $k$-means clustering \code{n\_cluster = 3}.

\begin{Schunk}
\begin{Sinput}
vm_res <- ce_estimate(y = data$y, x = data$covariates, w = data$w, method = "VM", 
                      estimand = "ATT", reference_trt = 1, n_cluster = 3)
\end{Sinput}
\end{Schunk}

The number of matched individuals is also stored in the output list:

\begin{Schunk}
\begin{Sinput}
vm_res$number_matched
\end{Sinput}
\begin{Soutput}
#> 158
\end{Soutput}
\end{Schunk}

\paragraph{Targeted maximum likelihood estimation}
TMLE is a doubly robust approach that combines outcome estimation, IPTW estimation, and a targeting step to optimize the parameter of interest with respect to bias and variance. \cite{rose2019double} implemented TMLE to estimate the ATE effects of multiple treatments. \pkg{CIMTx} calls the R package \CRANpkg{tmle} to implement TMLE for the ATE effects. As suggested by \cite{rose2019double}, nonparametric boostrap is used in \pkg{CIMTx} to obtain the confidence interval of the treatment effect estimate.

Calling \code{method = "TMLE"} implements TMLE in \pkg{CIMTx}. We use nonparametric bootstrap to estimate the 95\% confidence intervals by setting  \code{boot = TRUE} and \code{nboots = 100}.

\begin{Schunk}
\begin{Sinput}
tmle_res_boot <- ce_estimate(y = data$y, x = data$covariates, w = data$w, nboots = 100, 
                             method = "TMLE", estimand = "ATE", boot = TRUE, 
                             sl_library = c("SL.glm", "SL.glmnet", "SL.rpart"))
\end{Sinput}
\end{Schunk}

\begin{Schunk}
\begin{Sinput}
summary(tmle_res)
\end{Sinput}
\begin{Soutput}
#> $ATE12
#> EST   SE LOWER UPPER
#> RD -0.36 0.04 -0.45 -0.29
#> RR  0.30 0.05  0.21  0.39
#> OR  0.17 0.04  0.11  0.24

#> $ATE13
#> EST   SE LOWER UPPER
#> RD -0.60 0.04 -0.67 -0.51
#> RR  0.20 0.03  0.15  0.28
#> OR  0.06 0.02  0.04  0.10

#> $ATE23
#> EST   SE LOWER UPPER
#> RD -0.24 0.05 -0.34 -0.14
#> RR  0.68 0.06  0.57  0.79
#> OR  0.34 0.09  0.21  0.55
\end{Soutput}
\end{Schunk}

\subsection{Identification of a common support region} \label{sec:positivity}
Turning to causal identification assumptions. If the positivity assumption (A2) is violated, problems can arise when extrapolating over the areas of the covariate space where common support does not exist. It is important to define a common support region to which the causal conclusions can be generalized. In \pkg{CIMTx}, the identification of a common support region is offered in three methods: IPTW, VM and BART. 

\begin{figure}[H]
\centering
\includegraphics[width = 0.95\textwidth]{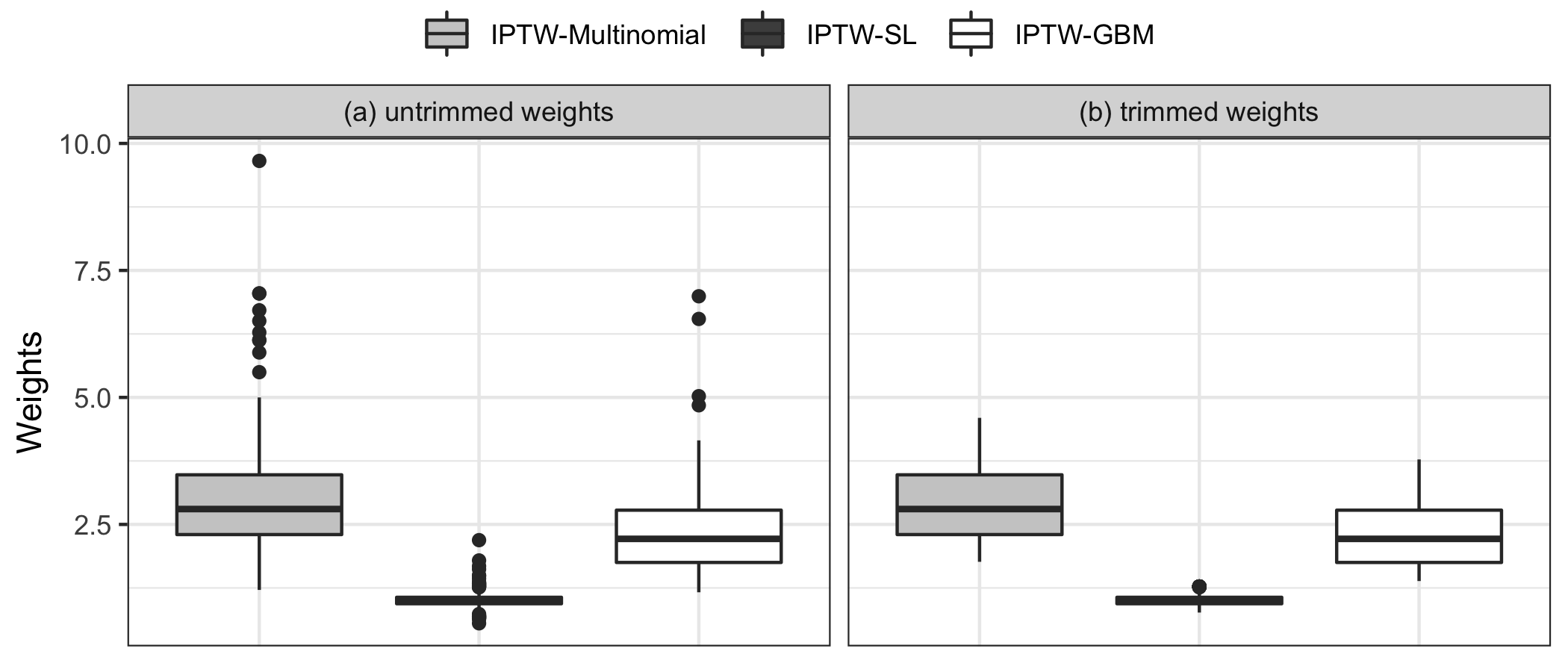}
\caption{Distributions of the inverse probability of treatment weights estimated by multinomial logistic regression, super learner and generalized boosted models. Panel (a) shows results before weight trimming. Panel (b) displays results after trimming the weights at 5\% and 95\% of the distribution. }
\label{fig:p_est_weights}
\end{figure}

For IPTW, one strategy is weight truncation, by which extreme weights that fall outside a specified range limit of the weight distribution are set to the range limit. This functionality is offered in  \pkg{CIMTx} via the \code{trim\_perc} argument. \code{trim\_perc}, which can take two values -- one for the lower- and one for the upper-percentile of the weight distribution for trimming. Figure~\ref{fig:p_est_weights} shows the distributions of the weights estimated by the three methods before and after weight trimming at the 5\% and 95\% of the weight distribution. 

\begin{Schunk}
\begin{Sinput}
plot(iptw_multi_res, iptw_sl_res, iptw_gbm_res, iptw_multi_trim_res, 
     iptw_sl_trim_res, iptw_gbm_trim_res)
\end{Sinput}
\end{Schunk}

For VM,  \cite{lopez2017estimation} proposed a rectangular support region defined by the maximum value of the smallest GPS and the minimum value of the largest GPS among the treatment groups. Individuals that fall outside the region are discarded from the causal analysis. This feature is automatically implemented with \code{"VM"} in \pkg{CIMTx}. 

For BART, \cite{hu2020estimation} supplied BART with a strategy to identify a common support region for retaining inferential units, which is to discard individuals with a large variability in their predicted potential outcomes. Specifically, for the ATT effects, any individual $i$ with $W_i = w$ will be discarded if
\begin{equation} \label{eq:discard}
  s_i^{f_{w^\prime}} > \text{max}_j \{s_j^{f_w} \}, \forall j: W_j = w, w^\prime \neq w \in \mathscr{W},   
\end{equation}
 where $s_j^{f_w}$ and $s_i^{f_{w^\prime}}$ respectively denote the standard deviation of the posterior distribution of the potential outcomes under treatment $W = w$ and $W = w^\prime$, for a given sample $j$. For the ATE effects, the discarding rule in equation~\eqref{eq:discard} is applied to each treatment group. Users can implement the discarding rule by setting the \code{discard} argument in \pkg{CIMTx}. Using $ATT_{1|1,2}$ as an example, 5 (\code{bart\_dis\_res\$n\_discard}) individuals in the reference group $w=1$ were discarded from the simulated data. 

\begin{Schunk}
\begin{Sinput}
bart_dis_res <- ce_estimate(y = data$y, x = data$covariates, w = data$w, 
                            method = "BART", estimand = "ATT", discard = TRUE, 
                            ndpost = 100, reference_trt = 1)
\end{Sinput}
\end{Schunk}



\subsection{Sensitivity analysis for unmeasured confounding}\label{sec:sa}
The violation of the ignorability assumption (A3) can lead to biased treatment effect estimates. Sensitivity analysis is useful in gauging how much the causal conclusions will be altered in response to different magnitude of departure from the ignorability assumption. \pkg{CIMTx} implements a new flexible sensitivity analysis approach developed by \cite{hu2022flexible}. This approach first defines a confounding function for any pair of treatments $(w, w')$ as
\begin{eqnarray} \label{eq:cf}
c(w, w', \bm{x}) = E \lsq Y(w) \cond W = w, \bm{X}=\bm{x}\rsq - E \lsq Y (w) \cond W =w', \bm{X}=\bm{x} \rsq.
\end{eqnarray} 

The  confounding function, also viewed as the sensitivity parameter in a sensitivity analysis, directly represents the difference in the mean potential outcomes $Y(w)$ between those treated with $W=w$ and those treated with $W=w'$, who have the same level of $\bm{x}$.  If the ignorability assumption holds, the confounding function will be zero for all $w \in \mathscr{W}$. When treatment assignment is not ignorable, the unmeasured confounding is present and the causal effect estimates using measured $\bm{X}$ will be biased. \cite{hu2022flexible} derived the form of the resultant bias as: 
\begin{align} \label{eq:biasform}
\begin{split}
\text{Bias}(w,w') =&-p_{w} c(w', w, \bm{x}) + p_{w'}c(w,w',\bm{x})\\
&-\sum\limits_{l: l \in \mathscr{W}\setminus\{w, w'\}} p_{l} \lbc c(w', l, \bm{x}) -c(w,l,\bm{x}) \rbc, 
\end{split}
\end{align}
where $p_{w} = P(W= w \cond \bm{X}= \bm{x})$, $w \neq w' \in \mathscr{W} = \{1, \ldots, T\} $.

Table~\ref{tab:cfun-inter} demonstrates the plausible assumptions about the confounding functions and their interpretations.
 There are three ways in which we can specify the prior for the confounding functions: (i) point mass prior; (ii) re-analysis over a range of point mass priors (tipping point); (iii) full prior with uncertainty specified. Since the new sensitivity analysis approach was developed within the Bayesian framework, strategy (iii) offers an advantage of incorporating the statistical uncertainty due to sampling and the uncertainty about the values of the sensitivity parameters. In strategy (i), a fixed value is assumed for the sensitivity parameter. Strategy (ii) expands on strategy (i) and  examines how the causal conclusion would change when a range of values are assumed for the sensitivity parameter. We will demonstrate all three cases of prior specifications with \code{sa()} function in \pkg{CIMTx} package. \cite{hu2022flexible} further discussed (a) strategies to specify the confounding functions that represent our prior beliefs about the degrees of unmeasured confounding via the remaining variability in the outcomes unexplained by measured $\bm{X}$ \citep{hogan2014bayesian}; and (b) ways in which the causal effects can be estimated adjusting for the presumed degree of unmeasured confounding. 

\begin{table}[H]
\centering
\caption{Interpretation of assumed priors on $c(w, w', \bm{x})$ and $c(w', w, \bm{x})$ for causal estimands based on the risk difference, assuming the outcome is an adverse event.}
\label{tab:cfun-inter}
\begin{tabular}{ccp{0.72\textwidth}}
\toprule
\multicolumn{2}{c}{Prior assumption}  & Interpretation and implications of the assumptions\\
$c(w, w', \bm{x})$ &$c(w', w, \bm{x})$ &\\\hline
$>0$ &$<0$ & Unhealthier individuals are treated with $w$. \\
$<0$ &$>0$ & Contrary to the above interpretation, unhealthier individuals are treated with $w'$. \\
$<0$ &$<0$ & The observed treatment allocation between $w'$ and $w$ is beneficial relative to the alternative which reverses treatment assignment for everyone. \\
$>0$ &$>0$ & Contrary to the above interpretation, the observed treatment allocation between $w'$ and $w$ is undesirable relative to the alternative which reverses treatment assignment for everyone.\\ 
\bottomrule
\end{tabular}
\end{table}

The proposed sensitivity analysis algorithm  proceeds with the following steps \citep{hu2022flexible}: 
\begin{enumerate}
    \item Fit a multinomial probit BART model \citep{kindo2016multinomial} $f^{\text{MBART}}(A\cond X)$ to estimate the GPS, $p_l \equiv P (W = l \cond X =x) \;  \forall l \in \mathscr{W}$, for each individual. 
    \item 
\begin{algorithmic}
\For{$w \gets 1$ to $T$}                    
 \State {Draw $M_1$ GPS $\tilde{p}_{l1}, \ldots, \tilde{p}_{lM_1}, \forall l \neq w \wedge l \in \mathscr{W}$ from the posterior predictive distribution of $f^{\text{MBART}} (W \cond \bm{X})$ for each individual.} 
    \For{$m \gets 1$ to $M_1$}                    
        \State {Draw $M_2$ values $\eta^*_{lm1}, \ldots, \eta^*_{lmM_2}$ from the prior distribution of each of the confounding functions $c(w, l, \bm{x})$, for each $l \neq j \wedge l \in \mathscr{W}$. } 
    \EndFor
    \EndFor
\end{algorithmic}
\item Compute the adjusted outcomes, $\YCF_i \equiv Y_i - \sum_{l \neq w}^T P(W_i = l\cond \bm{X}_i= \bm{x}) c(w, l,\bm{x})$,  for each treatment $w$,  for each of $M_1M_2$ draws of $\{\tilde{p}_{l1},  \eta^*_{l11}, \ldots, \eta^*_{l1M_2}, \ldots,  \tilde{p}_{lM_1},  \eta^*_{lM_11}, \ldots,  \eta^*_{lM_1M_2};  l \neq w \wedge l \in \mathscr{W}\}$. 
\item Fit a BART model to each of $M_1\times M_2$ sets of observed data with  the adjusted outcomes $\YCF$. 
\item Estimate the combined adjusted causal effects and uncertainty intervals by pooling posterior samples across model fits arising from the $M_1 \times M_2$ data sets.    
\end{enumerate}

We now demonstrate the Monte Carlo sensitivity analysis approach for unmeasured confounding \citep{hu2022flexible}. We first simulate a small dataset in a simple causal inference setting. There are two binary confounders: $X_1$ is measured and $X_2$ is unmeasured. 

\begin{Schunk}
\begin{Sinput}
set.seed(111)
data_SA <- data_sim(
  sample_size = 100, n_trt = 3,
  x = c("rbinom(1, .5)",  # x1: measured confounder
        "rbinom(1, .4)"), # x2: unmeasured confounder
  lp_y = rep(".2*x1+2.3*x2", 3),# parallel response surfaces
  nlp_y = NULL,
  align = F, # w model is not the same as the y model
  lp_w = c("0.2 * x1 + 2.4 * x2",  # w = 1
           "-0.3 * x1 - 2.8 * x2"),# w = 2
  nlp_w = NULL,
  tau = c(-2, 0, 2),  delta = c(0, 0), psi = 1)
\end{Sinput}
\end{Schunk}
Next we implement the  sensitivity analysis algorithm step-by-step.
\begin{enumerate}

\item Estimate the GPS for each individual.  Specifically, we fit a multinomial probit BART model regressing treatment assignment on covariates, using \code{mbart2} function from \CRANpkg{BART} package.  We set the number of posterior draws for the GPS (\code{m1}) to 50.

\begin{Schunk}
\begin{Sinput}
m1 <- 50; sample_gap <- 10
w_model <- BART::mbart2(x.train = data_SA$covariates, y.train = data_SA$w,
                        ndpost = m1 * sample_gap) 
\end{Sinput}
\end{Schunk}
\item Then we draw the GPS for each individual from the fitted  multinomial probit BART model.

\begin{Schunk}
\begin{Sinput}
gps <- array(w_model$prob.train[seq(1, m1 * sample_gap, sample_gap),], 
             dim = c(m1,  # 1st dimension is M1
                     length(unique(data_SA$w)), # 2nd dimension is w
                     dim(data_SA$covariates)[1])) # 3rd dimension is sample size
dim(gps)
\end{Sinput}
\end{Schunk}

\begin{Schunk}
\begin{Soutput}
#> 50 3 100
\end{Soutput}
\end{Schunk}

The output of the posterior GPS is a three-dimensional array. The first dimension  is the number of posterior draws for the GPS ($M_1$). The second dimension is the number of treatment $W$, and the third dimension is the total sample size.

\item Specify the prior distributions  and the number of draws ($M_2$) for the confounding functions $c(w,w',\bm{x})$. In this illustrative simulation example, we use the true values of the confounding functions within each stratum of $x_1$. This represents the strategy (i) point mass prior. 

\begin{Schunk}
\begin{Sinput}
x1 <- data_SA$covariates[, 1, drop = F]
x2 <- data_SA$covariates[, 2, drop = F] # x2 as the unmeasured confounder
w <- data_SA$w
x1w_data <- cbind(x1, w)
Y1 <- data_SA$y_true[, 1]
Y2 <- data_SA$y_true[, 2]
Y3 <- data_SA$y_true[, 3]
y <- data_SA$y
# Calculate the true confounding functions within x1 = 1 stratum
c_1_x1_1 <- mean(Y1[w == 1 & x1 == 1]) - mean(Y1[w == 2 & x1 == 1]) # c(1,2)
c_2_x1_1 <- mean(Y2[w == 2 & x1 == 1]) - mean(Y2[w == 1 & x1 == 1]) # c(2,1)
c_3_x1_1 <- mean(Y2[w == 2 & x1 == 1]) - mean(Y2[w == 3 & x1 == 1]) # c(2,3)
c_4_x1_1 <- mean(Y1[w == 1 & x1 == 1]) - mean(Y1[w == 3 & x1 == 1]) # c(1,3)
c_5_x1_1 <- mean(Y3[w == 3 & x1 == 1]) - mean(Y3[w == 1 & x1 == 1]) # c(3,1)
c_6_x1_1 <- mean(Y3[w == 3 & x1 == 1]) - mean(Y3[w == 2 & x1 == 1]) # c(3,2)

c_x1_1 <- cbind(c_1_x1_1, c_2_x1_1, c_3_x1_1, c_4_x1_1, c_5_x1_1,
                c_6_x1_1)# True confounding functions among x1 = 1
\end{Sinput}
\end{Schunk}

The true values of the confounding functions within the stratum $x_1 = 0$ can be calculated in a similar way.
\begin{Schunk}
\begin{Sinput}
c_1_x1_0 <- mean(Y1[w == 1 & x1 == 0]) - mean(Y1[w == 2 & x1 == 0])# c(1,2)
c_2_x1_0 <- mean(Y2[w == 2 & x1 == 0]) - mean(Y2[w == 1 & x1 == 0])# c(2,1)
c_3_x1_0 <- mean(Y2[w == 2 & x1 == 0]) - mean(Y2[w == 3 & x1 == 0])# c(2,3)
c_4_x1_0 <- mean(Y1[w == 1 & x1 == 0]) - mean(Y1[w == 3 & x1 == 0])# c(1,3)
c_5_x1_0 <- mean(Y3[w == 3 & x1 == 0]) - mean(Y3[w == 1 & x1 == 0])# c(3,1)
c_6_x1_0 <- mean(Y3[w == 3 & x1 == 0]) - mean(Y3[w == 2 & x1 == 0])# c(3,2)
c_x1_0 <- cbind(c_1_x1_0, c_2_x1_0, c_3_x1_0, c_4_x1_0, c_5_x1_0, c_6_x1_0) 
\end{Sinput}
\end{Schunk}

The true values of the confounding functions within the stratum of $x_1$ can be calculated using the helper function \code{true\_c\_fun\_cal()} in our package.

\begin{Schunk}
\begin{Sinput}
true_c_fun <- true_c_fun_cal(x = x1, w = w)
\end{Sinput}
\end{Schunk}

\item Calculate the confounding function adjusted outcomes with the drawn values of GPS and  confounding functions.

\begin{Schunk}
\begin{Sinput}
i <- 1; j <- 1
ycf <- ifelse(
  x1w_data[, "w"] == 1 & x1 == 1,
  # w = 1, x1 = 1
  y - (c_x1_1[i, 1] * gps[j, 2, ] + c_x1_1[i, 4] * gps[j, 3, ]),
  ifelse(
    x1w_data[, "w"] == 1 & x1 == 0,
    # w = 1, x1 = 0
    y - (c_x1_0[i, 1] * gps[j, 2, ] + c_x1_0[i, 4] * gps[j, 3, ]),
    ifelse(
      x1w_data[, "w"] == 2 & x1 == 1,
      # w = 2, x1 = 1
      y - (c_x1_1[i, 2] * gps[j, 1, ] + c_x1_1[i, 3] * gps[j, 3, ]),
      ifelse(
        x1w_data[, "w"] == 2 & x1 == 0,
        # w = 2, x1 = 0
        y - (c_x1_0[i, 2] * gps[j, 1, ] + c_x1_0[i, 3] * gps[j, 3, ]),
        ifelse(
          x1w_data[, "w"] == 3 & x1 == 1,
          # w = 3, x1 = 1
          y - (c_x1_1[i, 5] * gps[j, 1, ] + c_x1_1[i, 6] * gps[j, 2, ]),
          # w = 3, x1 = 0
          y - (c_x1_0[i, 5] * gps[j, 1, ] + c_x1_0[i, 6] * gps[j, 2, ])
        )
      )
    )
  )
) 
\end{Sinput}
\end{Schunk}

\item Use the adjusted outcomes to estimate the causal effects.

\begin{Schunk}
\begin{Sinput}
bart_mod_sa <- BART::wbart(x.train = x1w_data, y.train = ycf, ndpost = 1000)
predict_1_ate_sa <- BART::pwbart(cbind(x1, w = 1), bart_mod_sa$treedraws)
predict_2_ate_sa <- BART::pwbart(cbind(x1, w = 2), bart_mod_sa$treedraws)
predict_3_ate_sa <- BART::pwbart(cbind(x1, w = 3), bart_mod_sa$treedraws)
RD_ate_12_sa <- rowMeans(predict_1_ate_sa - predict_2_ate_sa)
RD_ate_23_sa <- rowMeans(predict_2_ate_sa - predict_3_ate_sa)
RD_ate_13_sa <- rowMeans(predict_1_ate_sa - predict_3_ate_sa)
predict_1_att_sa <- BART::pwbart(cbind(x1[w == 1,], w = 1), bart_mod_sa$treedraws)
predict_2_att_sa <- BART::pwbart(cbind(x1[w == 1,], w = 2), bart_mod_sa$treedraws)
RD_att_12_sa <- rowMeans(predict_1_att_sa - predict_2_att_sa) # w=1 is the reference
\end{Sinput}
\end{Schunk}

\end{enumerate}

Repeat steps 3 and 4  $M_1 \times  M_2$ times to form $M_1 \times M_2$ datasets with adjusted outcomes. The uncertainty intervals are estimated by pooling the posteriors  across the $M_1 \times M_2$ model fits. 

The \code{sa()} function implements the sensitivity analysis approach while fitting the $M_1 \times M_2$ models using parallel computation. 
\begin{Schunk}
\begin{Sinput}
sa_res <- sa(m1 = 50, m2 = 1, x = x1, y = data_SA$y, w = data_SA$w, ndpost = 100, 
             estimand = "ATE", prior_c_function =  true_c_fun, nCores = 3)
\end{Sinput}
\end{Schunk}

\begin{Schunk}
\begin{Sinput}
summary(sa_res)
\end{Sinput}
\begin{Soutput}
#>            EST   SE LOWER UPPER
#> ATE_RD12 -0.44 0.10 -0.64 -0.23
#> ATE_RD13 -0.58 0.11 -0.80 -0.36
#> ATE_RD23 -0.14 0.12 -0.38  0.10
\end{Soutput}
\end{Schunk}
We compare the sensitivity analysis results to the naive estimators where we ignore the unmeasured confounder $X_2$, and to the results where we had access to $X_2$. 
\begin{Schunk}
\begin{Sinput}
bart_with_x2_res <- ce_estimate(y = data_SA$y, x = cbind(x1, x2), w = data_SA$w, 
                                method = "BART", estimand = "ATE", ndpost = 100)
bart_without_x2_res <- ce_estimate(y = data_SA$y, x = x1, w = data_SA$w, 
                                   method = "BART", estimand = "ATE", ndpost = 100)
\end{Sinput}
\end{Schunk}

\begin{figure}[htbp]
\centering
\includegraphics[width = 1\textwidth]{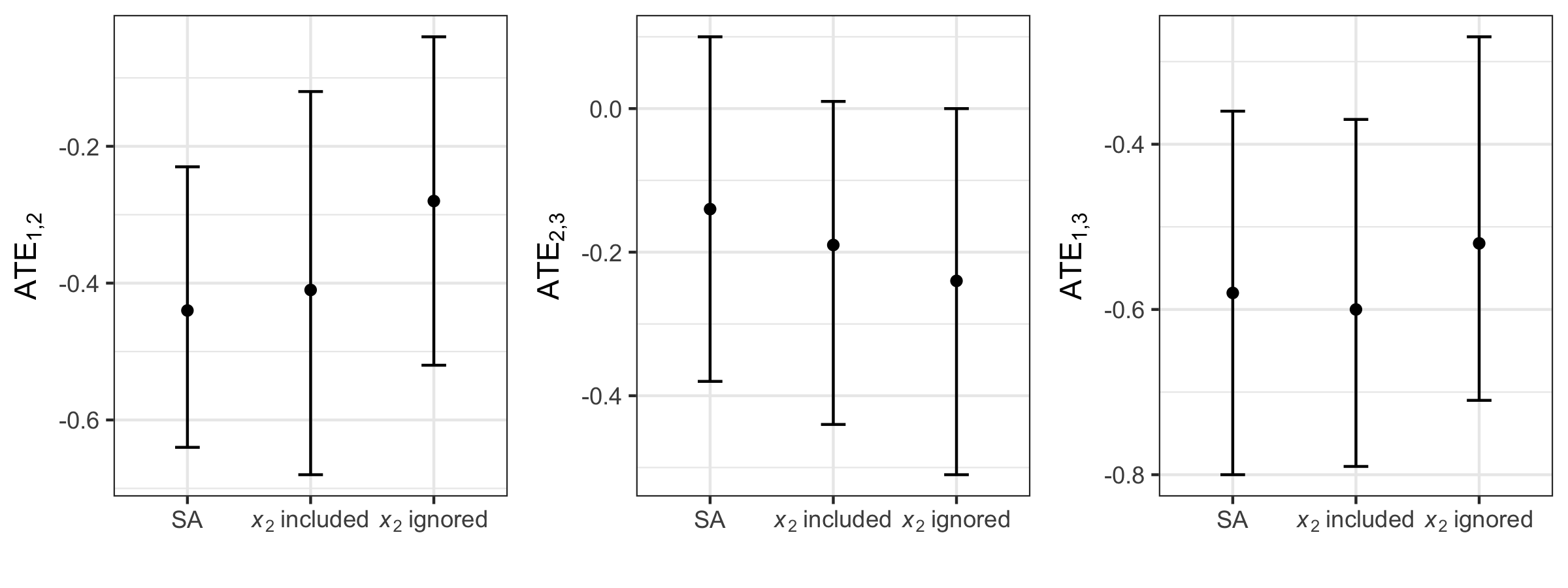}
\caption{Estimates and 95\% credible intervals for $ATE_{1,2}$, $ATE_{2,3}$ and $ATE_{1,3}$ }
\label{fig:p_SA_demo}
\end{figure}

Figure \ref{fig:p_SA_demo} compares the estimates of  $ATE_{1,2}$, $ATE_{2,3}$ and $ATE_{1,3}$ from the three analyses. The sensitivity analysis estimators are similar to the results that could be achieved had the unmeasured confounder $X_2$ been made available. 

We can also conduct the sensitivity analysis for the ATT effects by setting  \code{estimand = "ATT"}. 

\begin{Schunk}
\begin{Sinput}
sa_att_res <- sa(m1 = 50, m2 = 1, x = x1, y = data_SA$y, w = data_SA$w, ndpost = 100,
                 estimand = "ATT", prior_c_function =  true_c_fun, nCores = 1, 
                 reference_trt = 1)
\end{Sinput}
\end{Schunk}

\begin{Schunk}
\begin{Sinput}
summary(sa_att_res)
\end{Sinput}
\begin{Soutput}
#>            EST   SE LOWER UPPER
#> ATT_RD12 -0.42 0.10 -0.63 -0.22
#> ATT_RD13 -0.57 0.11 -0.79 -0.35
\end{Soutput}
\end{Schunk}

Finally, we demonstrate the \code{sa()} function in a more complex data setting with 3 measured confounders and 2 unmeasured confounders.  

\begin{Schunk}
\begin{Sinput}
set.seed(1)
data_SA_2 <- data_sim(
  sample_size = 100, n_trt = 3,
  x = c( "rnorm(0, 0.5)",  # x1
         "rbeta(2, .4)",   # x2
         "runif(0, 0.5)",  # x3
         "rweibull(1, 2)", # x4 as one of the unmeasured confounders
         "rbinom(1, .4)" ),  # x5 as one of the unmeasured confounders
  lp_y = rep(".2*x1 + .3*x2 - .1*x3 - 1.1*x4 - 1.2*x5", 3),
  nlp_y  = rep(".7*x1*x1  - .1*x2*x3", 3), # parallel response surfaces
  align = FALSE,
  lp_w =  c(".4*x1 + .1*x2 - 1.1*x4 + 1.1*x5",  # w = 1
            ".2*x1 + .2*x2 - 1.2*x4 - 1.3*x5"), # w = 2,
  nlp_w = c("-.5*x1*x4 - .1*x2*x5",  # w = 1
            "-.3*x1*x4 + .2*x2*x5"), # w = 2,
  tau = c(0.5,-0.5,0.5), delta = c(0.5,0.5), psi = 2)
\end{Sinput}
\end{Schunk}

We have demonstrated the strategy (i) point mass prior, and now show how strategy (ii) re-analysis over a range of point mass priors and (iii) full prior with uncertainty specified can be used. For strategy (ii), we can specify the grid of the specific confounding function using \code{seq()} function. In the following example, we will set  $c(1,3)$  as a grid of 5 negative numbers from -0.6 to 0 with an increment of 0.15, and set  $c(3,1)$ as a grid of 5 positive numbers from  0 to 0.6 with an increment of 0.15. This specification encodes our belief that unhealthier (suppose the outcome is death) individuals are treated with treatment option 3 (see Table~\ref{tab:cfun-inter})  because those receiving $w=1$ would have lower probability of death to either treatment.  The other confounding functions are drawn from a uniform distribution (strategy [iii]).


\begin{Schunk}
\begin{Sinput}
c_grid <- c("runif(-.6, 0)", "runif(0,.6)", "runif(-.6,0)", # c(1,2), c(2,1), c(2,3)
            "seq(-.6, 0,.15)", "seq(0,.6,.15)", "runif(0,.6)") # c(1,3), c(3,1), c(3,2)
SA_grid_res <- sa(y = data_SA_2$y, w = data_SA_2$w, x = data_SA_2$covariates[,-c(4,5)],
                  prior_c_function = c_grid, m1 = 1, nCores = 3, estimand = "ATE")
\end{Sinput}
\end{Schunk}

\begin{figure}[htbp]
\centering
\includegraphics[width = 0.65\textwidth]{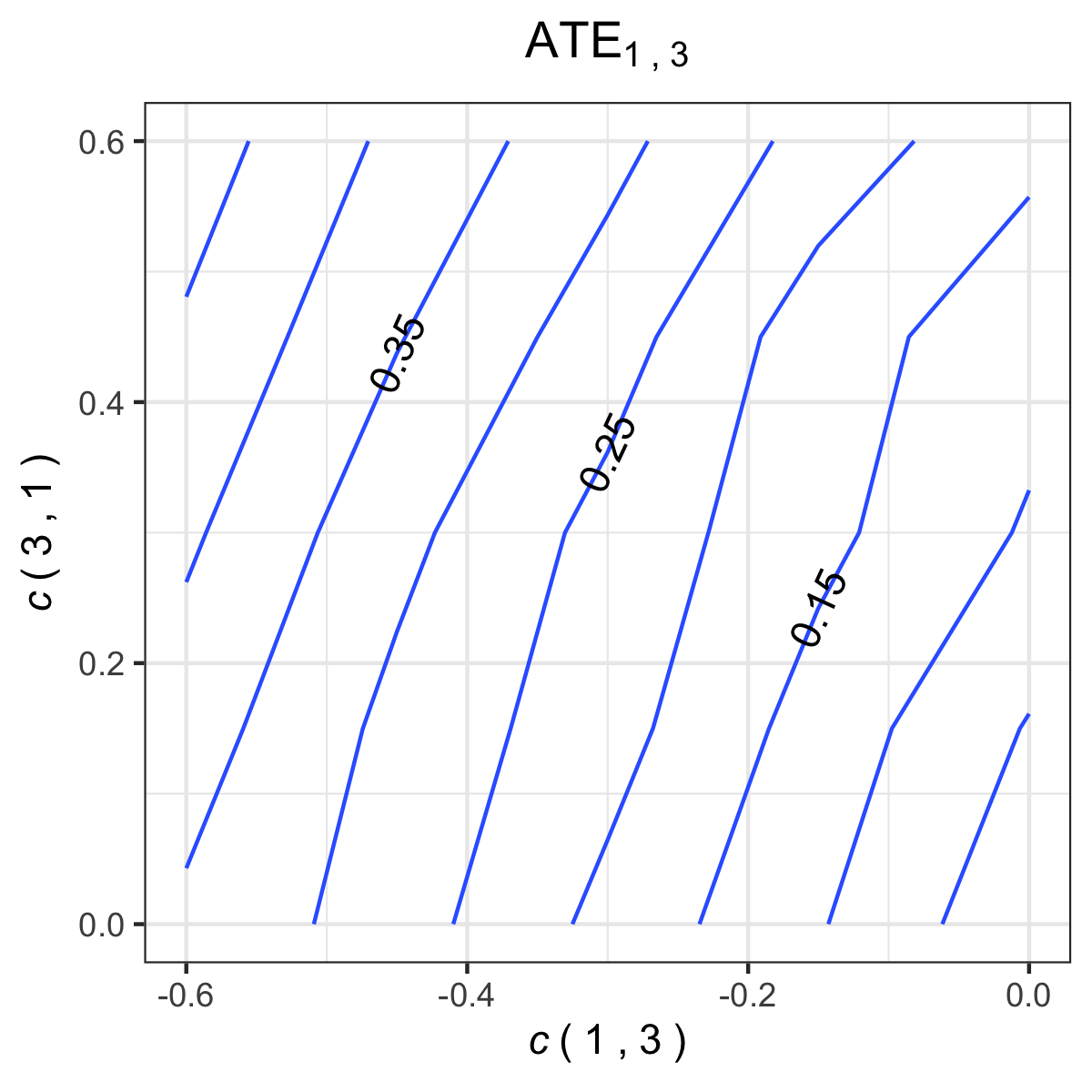}
\caption{Contour plot of the confounding function adjusted $ATE_{1,3}$. The blue lines report the adjusted causal effect estimates corresponding to pairs of values for $c(1,3)$ and $c(3,1)$ spaced on a grid $(-0.6, 0) \times (0, 0.6)$ incremented by 0.15, and under the prior distributions: $c(1,2), c(2,3) \sim U(-0.6,0);  c(2,1), c(2,3) \sim U(0, 0.6)$. }
\label{fig:p_contour}
\end{figure}

The sensitivity analysis results can be visualized via a contour plot. Figure \ref{fig:p_contour} shows how the estimate of $ATE_{1,3}$ would change under different 
 values of the two confounding functions $c(1,3,\bm{x})$ and $c(3,1,\bm{x})$.   Under the assumption that unhealthier patients are treated with $w=3$, when the effect of unmeasured confounding increases (moving up along the $-45^{\circ}$ line), the beneficial treatment effect associated with $w=3$ becomes more pronounced evidenced by larger estimates of $ATE_{1,3}$. 

\begin{Schunk}
\begin{Sinput}
plot(SA_grid_res, ATE = "1,3")
\end{Sinput}
\end{Schunk}

\section{Discussion}
 We contribute a comprehensive R package \pkg{CIMTx} suitable for causal analysis of observational data with multiple treatments and a binary outcome. In this package, we introduce six methods for the estimation of causal effects, including both the classical approaches and machine learning based methods. Drawing causal inference from non-experimental data inevitably involves structrual causal assumptions. \pkg{CIMTx} offers a unique set of features to address two key assumptions: positivity and ignorability, using appropriate estimation procedures. Additionally,  the \pkg{CIMTx} package provides guidance to readers on how to simulate data possessing the data characteristics in the multiple treatment setting. Detailed step-by-step examples are provided to demonstrate all methods.  The current version of the \pkg{CIMTx} package focuses on binary outcomes. For future research, developing methods and R packages for causal inferences with more complex outcomes such as censored survival outcomes \citep{hu2022aflexible} could be a worthwhile contribution. 
 
\bibliography{CIMTx}

\address{Liangyuan Hu\\
  Department of Biostatistics and Epidemiology\\
  Rutgers University School of Public Health\\
  683 Hoes Lane West\\
  Piscataway, NJ 08854, United States of America\\
  }
\email{liangyuan.hu@rutgers.edu}

\address{Jiayi Ji\\
  Department of Biostatistics and Epidemiology\\
  Rutgers University School of Public Health\\
683 Hoes Lane West\\
  Piscataway, NJ 08854, United States of America\\
}
\email{jj869@sph.rutgers.edu}

\end{article}

\end{document}